\documentstyle[12pt,epsf]{article}

\newcommand{\ewxy}[2]{\setlength{\epsfxsize}{#2}\epsfbox[10 60 640 570]{#1}}

\newtheorem{statement}{Statement}

\begin{document}

\begin{center}{\Large\bf Instantons and Chiral Symmetry on the
Lattice}\\[2cm] 
{\bf Tam\'as G. Kov\'acs} \\
{\em Instituut-Lorentz for Theoretical Physics, P.O.Box 9506 \\
2300 RA, Leiden, The Netherlands}\\
{\sf e-mail: kovacs@lorentz.leidenuniv.nl}\\[15mm] 
\end{center}

\begin{abstract}
I address the question of how much of QCD in the chiral 
limit is reproduced by instantons. After reconstructing 
the instanton content of smoothed Monte Carlo lattice 
configurations, I compare hadron spectroscopy
on this instanton ensemble to the spectroscopy on  the original
``physical'' smoothed configurations using a chirally 
optimised clover fermion action. By studying the zero mode
zone in simple instances I find that the optimised action
gives a satisfactory description of it. Through the Banks-Casher
formula, instantons by themselves are shown to break 
chiral symmetry but hadron correlators on the instanton
backgrounds are strongly influenced by free quark propagation.
This results in unnaturally light hadrons and a small splitting 
between the vector and the pseudoscalar meson channels.
Superimposing a perturbative ensemble of
zero momentum gauge field fluctuations (torons) on the instantons
is found to be enough to eliminate the free quarks 
and restore the physical hadron correlators. I argue that the 
torons that are present only in finite volumes, are probably
needed to compensate the unnaturally large finite size effects
due to the lack of confinement in the instanton ensemble.
\end{abstract}

\section{Introduction}

According to the instanton liquid model (ILM) instantons are largely
responsible for the spontaneous breaking of chiral symmetry 
and the low energy properties of light hadrons \cite{Shuryak}.
This has been demonstrated using phenomenological instanton
liquid models. The two main tasks involved in the construction are
\begin{itemize}
\item{Producing an equilibrium instanton ensemble assuming
      some interaction between the pseudoparticles. This is 
      typically done with a variational calculation \cite{Diakonov}.}
\item{Computing hadronic observables using the above instanton
      ensemble as the gauge background.}
\end{itemize}
This is a formidable task and the computation necessarily involves
several approximations. It would thus be very desirable to check
the physical picture emerging from the ILM, starting from 
first principles.

In the last few years there has been a considerable effort 
in this direction, using lattice simulations of QCD as a tool.
The properties of the instanton ensemble have been extracted
from Monte Carlo generated lattice configurations using 
several different ``smoothing'' techniques. \cite{everybody}.
The most important parameters, the topological 
susceptibility, the instanton density, 
and the average instanton size were found to be in qualitative
agreement with the ILM. On the other hand --- although there 
is a considerable amount of indirect evidence 
\cite{indirect,For_ch,Negele} --- so far the lattice has not given 
any direct results concerning the second part of the 
above programme. 

In Ref. \cite{Negele} cooled lattice configurations were used 
to show that low eigenmodes of the Dirac operator saturate the
point-to-point hadronic correlators. More recently a strong 
correlation was found between the chiral density of low modes
of the staggered Dirac operator and the topological charge density
\cite{For_ch}. These results suggest that instantons
probably really play an important role in QCD. All this work
however has involved Monte Carlo generated or 
cooled lattices that contained other gauge field fluctuations
in addition to the instantons. Therefore, based solely on these studies
it is impossible to separate the part of the effects due to the 
instantons and the part (if any) that is due to the rest of the gauge field 
fluctuations. 

To my knowledge, the only lattice study so 
far that makes such a separation possible has 
appeared in Ref. \cite{spect}. There, the instanton
content of smoothed gauge configurations was reconstructed. Hadron
spectroscopy done on these artificial instanton backgrounds was 
compared to the spectroscopy on the corresponding smoothed 
configurations. While the smoothed ensemble yielded physically
reasonable results for the pion and the rho mass, the instantons
failed to reproduce this. In fact Wilson spectroscopy on the
instanton ensemble was hardly different from free field theory
(trivial gauge field background). In view of the positive
results cited above, this was highly unexpected, especially
that at first sight the only difference between the compared
two ensembles appeared to be the presence versus the lack of 
confinement on the smoothed and the artificial instanton ensembles 
respectively. According to the ILM, confinement is not essential 
for the low energy properties of light hadrons \cite{Shuryak}. 
The findings of Ref.\ \cite{Negele} indeed suggest that the
physics in the chiral limit is rather insensitive to the string
tension. With less than half of the original string tension,
the low fermion modes on the cooled configurations still 
produced physically sensible hadron correlators.

In the present paper I address the question of how to reconcile
the insensitivity of light hadrons to the string tension
(also expected from the ILM) with the failure of the instantons
alone to reproduce the correct physics (found in \cite{spect}).
I will exclusively consider the SU(2) gauge group.
Along the way I will also obtain some useful insight into
the structure of the low end of the spectrum of lattice Dirac operators
and in particular I will show the importance of using a (close to) chiral
lattice Dirac operator. This is essential for a faithful 
reproduction of continuum-like quark (quasi)zero modes. 

The plan of the paper is as follows. In Section \ref{se:zmz} I 
briefly discuss the instanton liquid picture of physics in the
chiral limit, and in particular the role of the zero modes 
and the so called ``zero mode zone''. In Section \ref{se:cs}
I point out the role of the explicit chiral symmetry
breaking of the lattice Dirac operator and the importance of
minimising it. I show that the optimised clover Dirac
operator introduced in \cite{exceptional} performs much better
in this respect than the Wilson operator. I explicitly study the 
zero mode zone in the simplest nontrivial case, that of an 
instanton antiinstanton pair and find that the mixing of zero
modes closely follows the pattern expected in the continuum.

In Section \ref{se:spectrum}
I find that the optimised clover operator produces
spectroscopy results markedly different from the Wilson operator
but it still falls short of reproducing the correct physics on
the instanton backgrounds. Instead, it yields a peculiar mixture
of QCD and free field theory. I demonstrate that this is due to
the presence of a large number of low lying free field modes 
in the spectrum of the Dirac operator with the instanton backgrounds 
that produce artificially light hadrons. The free modes are completely 
absent from the smoothed configurations.
They can be also eliminated from the instanton configurations
by superimposing an ensemble of zero momentum gauge field fluctuations
(torons) over the instantons. These gauge field fluctuations
survive even deep in the perturbative regime. I argue that 
they are probably needed to compensate for the anomalously large
small-volume effects present in the instanton ensemble due to
the lack of confinement. By comparing the density of eigenvalues 
around zero on the physical (smoothed), the toron, and toron
plus instanton ensembles, I obtain direct evidence that the 
instantons are responsible for chiral symmetry breaking.
After the elimination of the free quark
modes, the instanton backgrounds are shown to reproduce the 
correct physical pion and rho correlators. 

Section \ref{se:con}
contains my conclusions along with some speculations on 
still unanswered questions. Finally, in the Appendix, 
for reference I collected some general properties of Wilson 
type lattice Dirac operators that I used in the main text. 
Some of these can be also found in the literature.

\section{The Instanton Liquid Model and the Zero Mode Zone}
\label{se:zmz}

In this section I describe how instantons can produce
low lying modes of the Dirac operator and why this is 
important for the physics in the chiral limit.
For simplicity I start with the continuum 
theory and in the second part of this Section I 
discuss the complications that appear on the lattice.

\subsection{The Continuum}

Let $D$ be the continuum massless Dirac operator which 
implicitly depends on a gauge field background. The basic
building block of the physics in the chiral limit is the 
$m \rightarrow 0$ limit of the massive quark propagator,
$(D-m)^{-1}$. From the spectral decomposition in terms
of eigenmodes of $D$
\begin{equation}
  (D-m)^{-1} = \sum_i \frac{1}{\lambda_i-m} \;\; 
                             |i\rangle \langle i|,
\end{equation}
it is easily seen that in the chiral limit the most important 
modes are the ones with eigenvalues $\lambda_i$ close to 0.
The crucial assumption of the ILM is that it is the instantons
that are responsible for generating the bulk of these lowest eigenmodes.

This can be qualitatively understood as follows. According to the 
Atiyah-Singer index theorem, in the background of an instanton
(antiinstanton) $D$ has at least one negative (positive) 
chirality zero mode. In the presence of an infinitely separated
instanton antiinstanton pair we still expect to find 
opposite chirality zero modes localised on the two objects.
If the members of the pair are brought closer to one another,
the two degenerate zero eigenvalues will in general split into
two complex ones still close to the origin. A very simplistic but
qualitatively correct description of this can be given as 
follows. 

Let $\psi_I$ and $\psi_A$ be the zero modes of the Dirac operators
$D(I)$ and $D(A)$ respectively. Since we consider $D$ in different 
gauge backgrounds, we explicitly indicate this; $I$ and $A$ refer
to an instanton and an antiinstanton. We now want to describe
the spectrum of $D(IA)$ where $IA$ is some superposition
of the gauge fields $I$ and $A$. Since $\psi_I$ and $\psi_A$
are of opposite chirality, $\psi_I^\dagger \psi_A = 0$. 
Let us arbitrarily complete the set of these two vectors into an 
orthonormal basis and construct the matrix of $D(IA)$ in this basis. 
The elements of this matrix in the subspace spanned by
$\psi_I$ and $\psi_A$ are
\begin{equation}
\left( \begin{array}{cc}
          0           &       T \\
         -T           &       0
       \end{array}
\right),
    \label{eq:Dzmz}
\end{equation}
where 
\begin{equation}
      T \, = \, \psi_I^\dagger D(IA) \psi_A. 
\end{equation}
The diagonal matrix elements
vanish because $D$ maps left handed vectors into right handed ones
and vice versa. By a suitable choice of the phases, the off-diagonal
elements can always be made real and due to the anti-Hermiticity
of $D$ they are of equal magnitude and have opposite signs. 
If we now assume that $D(IA)\psi_I$ and 
$D(IA)\psi_A$ lie approximately in this two-dimensional subspace,
i.e. the matrix of $D(IA)$ contains the above  $2\times2$ 
block-diagonal part, then $D(IA)$ can be easily seen to have
two complex conjugate eigenvalues $\pm iT$ with the corresponding
(unnormalised) eigenvectors being $\psi_I \pm i  \psi_A$.

This mechanism can be generalised to gauge fields which are
superpositions of several instantons and antiinstantons. The
basis in this case consists of the zero modes of the 
individual (anti)instantons. The only additional approximation
involved is the assumption that the zero modes corresponding 
to different (anti)instantons are orthogonal. If the pseudoparticles
are well separated, this is approximately true. One can thus
consider the matrix of $D$ restricted to the subspace of zero modes ---
called the ``zero mode zone'' --- in a general instanton background.
If the off diagonal elements of this matrix are small then
it will have complex eigenvalues close to the origin, in addition 
to a number of exact zero eigenvalues corresponding to the total 
topological charge. Moreover, the complex eigenmodes will be
approximately linear combinations of several instanton and antiinstanton
zero modes and thus become highly delocalised, making it possible
to propagate quarks to large distances. 

The most important quantities in this construction are the
off diagonal instanton antiinstanton matrix elements, the $T$'s.
In the simplest case of only one pair, $T$ is known to
depend on the distance and the relative orientation
of the members of the pair as
\begin{equation}
  T = \mbox{tr}(g^\dagger \hat{R}) \; f(\rho_I,\rho_A,|R|),
     \label{eq:relor}
\end{equation}
where $g \in SU(2)$ describes the relative orientation in group space,
\begin{equation}
  \hat{R}=(R_0+iR_k\sigma_k)/|R|,
\end{equation}
and $R$ is the relative
position of the instanton and the antiinstanton \cite{Shuryak}. 
$f$ is a function of the instanton and antiinstanton scale
parameters and the distance between them. At small distances
it depends on the ansatz used to combine the field of the
instanton and the antiinstanton, at large separation it is 
expected to fall off as $1/|R|^3$.

\subsection{Complications Arising on the Lattice}

We have seen how instanton zero modes can give rise
to the low-lying eigenmodes that are the most important ones
in the chiral limit. Close to the chiral limit the number
of modes giving a substantial contribution to the quark 
propagator becomes smaller and smaller. It is thus crucial
to reproduce these low eigenmodes and eigenvalues as faithfully
as possible on the lattice if we want to test the ILM.

The most important obstruction to this is that due to the
Nielsen-Ninomiya theorem it is impossible to construct a local
lattice Dirac operator describing one fermion species 
with exact chiral symmetry \cite{Ni-Ni}. In the present paper
I shall limit the discussion to Wilson type lattice Dirac
operators that describe one fermion species but explicitly
break chiral symmetry. In the absence of
chiral symmetry the fermion zero modes corresponding to 
the topology of the gauge field are not protected; any
small fluctuation of the gauge field can shift the zero 
eigenvalues. 

In order to understand how this happens, a useful starting
point is the following property shared by the lattice and the continuum 
Dirac operator. Let $\psi_\lambda$ be an eigenvector of $D$ 
corresponding to the eigenvalue $\lambda$. If $D$ has no degeneracies 
then $\psi_\lambda^\dagger \gamma_5 \psi_\lambda \neq 0$ 
if and only if $\lambda$ is real. This is a simple consequence
of the so-called ``$\gamma_5$ Hermiticity'',
$D^\dagger = \gamma_5 \; D \; \gamma_5$,
a common property of the continuum and the lattice Dirac operator.
A proof of the above statement is given in the Appendix.

In the continuum, $D$ is anti-Hermitian and has eigenvalues
on the imaginary axis. Thus the only real eigenvalue it can have
is 0. Chirality makes a sharp distinction between zero and 
non-zero modes in the continuum. Zero modes have a chirality of
$\psi^\dagger \gamma_5 \psi=\pm 1$ while all 
the non-zero eigenmodes have zero chirality. 
Since the eigenmodes and their chiralities depend continuously
on the gauge field, zero modes are protected, they cannot be
shifted away from zero by continuous deformations of the gauge
field.

On the other hand, the lattice Dirac operator is not 
anti-Hermitian, and its eigenvalues are not constrained 
to be on the imaginary axis. While lattice zero modes are still
protected from being shifted off the real axis by smooth
deformations of the gauge field (this would cause their
chirality to jump to zero), there is nothing preventing them
from moving continuously along the real axis. Indeed,
a lattice Dirac operator will in general not have any 
exact zero modes unless the gauge field is fine tuned.
It is now clear that the lattice analogues of the continuum
zero modes are real modes of the Dirac operator. These 
have non-vanishing chiralities but in general their magnitude
is smaller than 1.

The crucial role chiral symmetry plays in this discussion
can be understood by noting that chiral symmetry of $D$
(i.e. $D\gamma_5=-\gamma_5D$) together with $\gamma_5$
Hermiticity would imply that $D$ is anti-Hermitian and
has protected zero modes. The smaller the explicit 
chiral symmetry breaking of $D$ is, the more its low
lying spectrum will resemble that of an anti-Hermitian
operator. Recently it has been shown that the closest a
lattice Dirac operator can come to being chirally symmetric
is by obeying the Ginsparg-Wilson relation and thus having an 
ultralocal chirality breaking in the quark propagator \cite{HP}.
In this case the spectrum of $D$ lies on a circle in the complex
plane passing through the origin. Indeed the 
low end of this spectrum is almost on the imaginary axis and 
there are protected zero modes.

I shall now discuss how the continuum description of the zero mode
zone presented in the previous Subsection has to be modified 
on the lattice. Assume that there is an instanton on the lattice.
In the limit when the instanton size goes to infinity
(in units of the lattice spacing), the instanton ``does not see''
the lattice at all, and the continuum description applies.
The lattice Dirac operator, $D(I)$ has an exact zero eigenvalue with 
a corresponding $-1$ chirality eigenmode $\psi_I$. If the
instanton is not infinitely large, the real eigenvalue 
gets shifted away from zero, its magnitude depends on
the instanton size and the chirality breaking of $D(I)$ 
\cite{exceptional,singleI}. The corresponding 
eigenmode also becomes only an approximate
eigenvector of $\gamma_5$ and thus $-1<\psi_I^\dagger \gamma_5
\psi_I<0$. 

Let us now consider the Dirac operator
in the presence of an instanton antiinstanton pair. Following the
continuum discussion in the previous Subsection, its matrix
elements between the (lattice approximate) ``zero modes'' $\psi_I$ and
$\psi_A$ can be parametrised as
\begin{equation}
\left( \begin{array}{cc}
        \psi^\dagger_I D(IA) \psi_I  &  \psi^\dagger_I D(IA) \psi_A\\
        \psi^\dagger_A D(IA) \psi_I  &  \psi^\dagger_A D(IA) \psi_A
       \end{array}
\right) \; = \;
\left( \begin{array}{cc}
         \mu+\delta       &       T_1 \\
         -T_2                 &   \mu-\delta,
       \end{array}
\right),
    \label{eq:Dzmzlat}
\end{equation}
where $\delta$ is real, $\lambda, T_1$ and $T_2$ are non-negative real
numbers. In the SU(2) case the diagonal elements are automatically
real and the off-diagonal ones can be also made real by a suitable
choice of the phases of the eigenvectors. This will be proved in
the Appendix. Due to the $\gamma_5$ Hermiticity of $D$, since $\psi_I$
and $\psi_A$ are approximate eigenvectors of $\gamma_5$, the 
off-diagonal terms have opposite signs and about the same
magnitude. The eigenvalues of this matrix are
\begin{equation}
   \lambda_{1,2} = \mu \pm i \sqrt{T_1 T_2-\delta^2},
\end{equation}
with the corresponding (unnormalised) eigenvectors being
\begin{equation}
    \psi_{1,2} = \frac{1}{T_2} \left( \sqrt{T_1 T_2-\delta^2} \;
                 \pm \; i \delta \right) \psi_I \, \pm \, i\psi_A.
\end{equation}
The most important feature of these eigenvectors is that 
--- as in the continuum --- they are 
mixtures of the instanton and antiinstanton zero modes with 
roughly the same magnitude (the ratio being $T_1/T_2$) and with 
a relative phase $\pi$, up to a correction proportional to 
$\delta$. This mechanism produces highly delocalised eigenmodes
in the zero mode zone and thus facilitates the propagation of quarks 
by jumping from instanton to antiinstanton. Maintaining this
feature on the lattice is thus crucial for the description 
of the zero mode zone. The requirement for the proper 
mixing of zero modes is the inequality
\begin{equation}
        T_1 T_2 > \delta^2.
  \label{eq:ineq} 
\end{equation}
For producing qualitatively continuum-like eigenmodes, the difference
of the diagonal elements has to be much smaller than the 
magnitude of the off-diagonal mixing matrix elements 
of $D(IA)$. Otherwise the two eigenvalues 
will be real and the corresponding eigenmodes rapidly become
localised on the instanton and the antiinstanton respectively.
Recall that in the continuum, the mixing matrix
elements $(T)$ are proportional to $\cos \phi$, where $\phi$
is the invariant angle of the instanton antiinstanton
relative orientation (see eq.\ (\ref{eq:relor})). $\delta$ (and $\mu$)
is expected to be of the order of the real eigenvalues of
$D(I)$ and $D(A)$, both small but nonzero numbers. This means that
inequality (\ref{eq:ineq}) will be violated for $\phi$ sufficiently 
close to $\pi/2$. We can minimise the range in relative 
orientation, where this happens by making $\delta$ small, i.e. having
the would-be zero modes close to zero.

\section{Chiral Symmetry and the Optimised Clover Operator}
\label{se:cs}

In this section I study how the general features of the zero mode zone
described in the previous Section are realised by two 
commonly used lattice Dirac operators, the Wilson and the clover
operator. The gauge background I use for this test contains an instanton
of size $1.5a$ and an antiinstanton of size $2.0a$, separated
by a distance $3.5a$ in the time direction. The lattice size
is $8^3\times16$ and I always use antiperiodic boundary conditions
in the time direction, periodic in all other directions. 
These parameters are typical in the spectroscopy calculations
that I shall present in the next Section.

Let me start with the Wilson operator. In the presence of
only the (anti)instanton with the above parameters, 
$D_w$ has a real eigenvalue at (0.17) 0.51 with chirality
(0.39) -0.53. It is clear that for these small instantons
the Wilson operator does not even come close to satisfying
the conditions of the previous Section: the ``zero modes''
are far away from zero with chiralities of magnitude much 
smaller than 1. The simplified description of the 
instanton-antiinstanton configuration in terms of the 
$2\times2$ matrix is then inadequate. If the instanton
and antiinstanton are close to being parallel $(\phi \approx 0)$
then the real part of the (anti)instanton related eigenvalues
is around 0.3. Moreover already at $\phi=\pi/4$ the two 
complex eigenvalues become real and the corresponding modes
are rather localised at the instanton and the antiinstanton
respectively.

A much better description of the zero mode 
zone is provided by the optimised
clover action proposed in Ref.\ \cite{exceptional}. 
(See also \cite{scaling} for more discussion and 
a recent scaling test of this action.) 
The clover coefficient was chosen to 
minimise the range in which the real
eigenvalues occur in the physical branch of the spectrum on
locally smooth gauge backgrounds. As explained in the previous
Section, this is exactly what is needed for a good description
of the mixing of zero modes. The optimal value for $c_{sw}$
was found to be around 1.2 on a set of APE smeared $\beta=5.7$ 
quenched SU(3) configurations. Physically these are quite similar
to the SU(2) configurations appearing in the present study, therefore
in what follows I shall always choose $c_{sw}=1.2$. 

\begin{figure}[!htb]
\begin{center}
\vskip 10mm
\leavevmode
\epsfxsize=120mm
\epsfbox{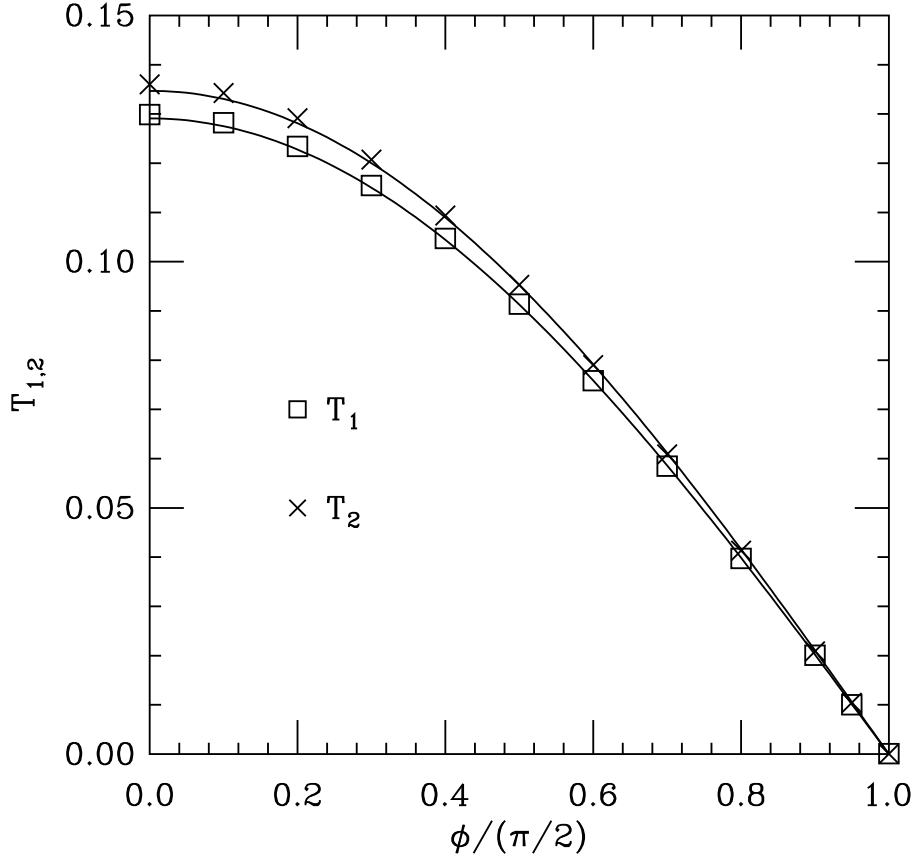}
\end{center}
\caption{The mixing matrix elements $T_1$ and $T_2$ versus the
invariant angle of the instanton-antiinstanton relative 
orientation. The solid lines are 1-parameter fits of the 
form const.$\times\cos \phi$, the form expected from the 
continuum computation.}
\label{fig:T12}
\end{figure}

The real modes occurring in the single (anti)instanton background
already reveal that the clover action is much closer to the 
continuum than the Wilson action. The eigenvalues are located at
-0.014 and -0.002 and the chiralities of the corresponding
eigenmodes are 0.993 and -0.995 respectively. In the instanton
plus antiinstanton background the lowest two modes are always
complex for almost any relative orientation of the pair, even
at $|\phi-\pi/2|=0.1$. We can hope that the simple description
in terms of the mixing matrix works well in this case. Indeed,
the magnitude of the diagonal elements is small, neither $\delta$
nor $\lambda$ goes above 0.02 in magnitude for any relative 
orientation. In Fig.\ \ref{fig:T12} I plotted the two off-diagonal
matrix elements,
\begin{equation}
  T_1 = \psi_I^\dagger D(IA) \psi_A \;\;
   \mbox{and} \;\; T_2 = |\psi_A^\dagger D(IA) \psi_I|,
\end{equation}
versus $\phi$ along with the one-parameter fits of the form 
const.$\times\cos \phi$. As expected, the mixing matrix elements
are very well described by the continuum ansatz of eq.\
(\ref{eq:relor}). A comparison of the eigenvalues of the
$2\times2$ mixing matrix and the explicitly computed eigenvalues 
of the lattice Dirac operator in the corresponding gauge backgrounds 
shows agreement to within $5\%$ in the whole range of $0<\phi<\pi/2$. 

We can conclude that the $c_{sw}=1.2$ clover operator should 
give a satisfactory treatment of the zero mode zone as long as 
the instanton size does not drop below around $1.5a$. 
This is of course true only if the gauge configuration
is locally smooth. Otherwise the fermion mass acquires 
an additive renormalisation which completely destroys
the above features important for the description of the zero 
mode zone.

\section{Hadron Spectroscopy and the Low Eigenmodes of the Dirac 
Operator }
\label{se:spectrum}

\subsection{Hadron Spectrum in Instanton Backgrounds}

Having a fermion action that is expected to describe
the zero modes and their mixing reasonably well, we can 
discuss the main topic of the paper, namely, what part of
QCD is reproduced by instantons alone. In this
Section I present quenched hadron spectroscopy results
obtained with the $c_{sw}=1.2$ clover fermion action.
The gauge field backgrounds on which the spectroscopy
is done are the ones used in Ref.\ \cite{spect}. The starting 
point is a set of 28 $8^3\times16$ SU(2) configurations generated
with a fixed point action. The lattice spacing is $a=0.145$fm,
as fixed by the Sommer parameter of the heavy quark potential.
These Monte Carlo generated configurations are not yet suitable
for our purposes since they are not locally ``smooth''. Both
the instantons and the details of the zero mode zone are
obscured by short distance fluctuations.

\begin{figure}[!htb]
\begin{center}
\vskip 10mm
\leavevmode
\epsfxsize=120mm
\epsfbox{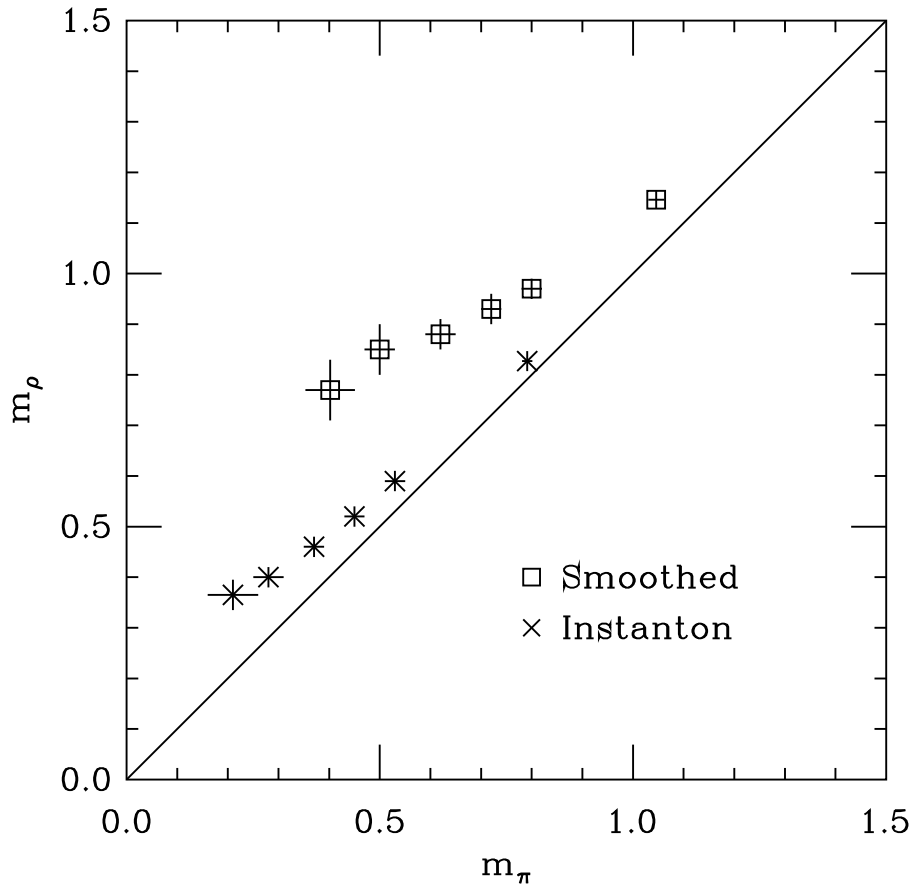}
\end{center}
\caption{$m_\pi$ versus $m_\rho$ measured on the smoothed
(boxes) and the instanton (crosses) backgrounds.}
\label{fig:pirho}
\end{figure}                               

Cycling, a smoothing technique based on the renormalisation
group, has been shown to preserve the main physical features
of the gauge configurations. After 9 cycling steps the
string tension \cite{Boulder1} and $m_\pi$ versus $m_\rho$
are essentially unchanged. Moreover, these lattices are smooth
enough that their instanton content can be unambiguously 
identified. In fact, about $70\%$ of the action is accounted
for by instantons (assuming there is no interaction between 
them). We can create artificial lattice configurations
that have the same instanton content as the smoothed ones.
The instanton sizes and locations are reproduced but 
the relative orientation in group space is distributed 
according to the SU(2) Haar measure. Recently a study of the 
relative orientation appeared in \cite{Ilgenfritz}, however 
reproducing the relative orientations would make our procedure
much more cumbersome.

I compare the spectroscopy done on the smoothed (cycled)
lattices and the corresponding artificial instanton configurations
with the same instanton content.
These ensembles are both locally smooth enough that
the optimised clover action should give a satisfactory 
description of their zero mode zone.
In Fig.\ \ref{fig:pirho} the vector meson mass is plotted
as a function of the pseudoscalar mass 
for both ensembles. The correlators were always fitted 
with the assumption that there is only
one lightest particle dominating both the pseudoscalar and
the vector channels. I also show
the $m_\pi=m_\rho$ free field line. In fact, the Wilson
fermion action produces exactly degenerate free field like
pions and rhos on the instanton backgrounds. Compared to the 
Wilson action, the clover action shows a marked improvement,
but it still fails to reproduce the $m_\pi$ vs.\ $m_\rho$ 
observed on the smoothed lattices. 
\begin{figure}[!htb]
\begin{center}
\vskip 10mm
\leavevmode
\epsfxsize=120mm
\epsfbox{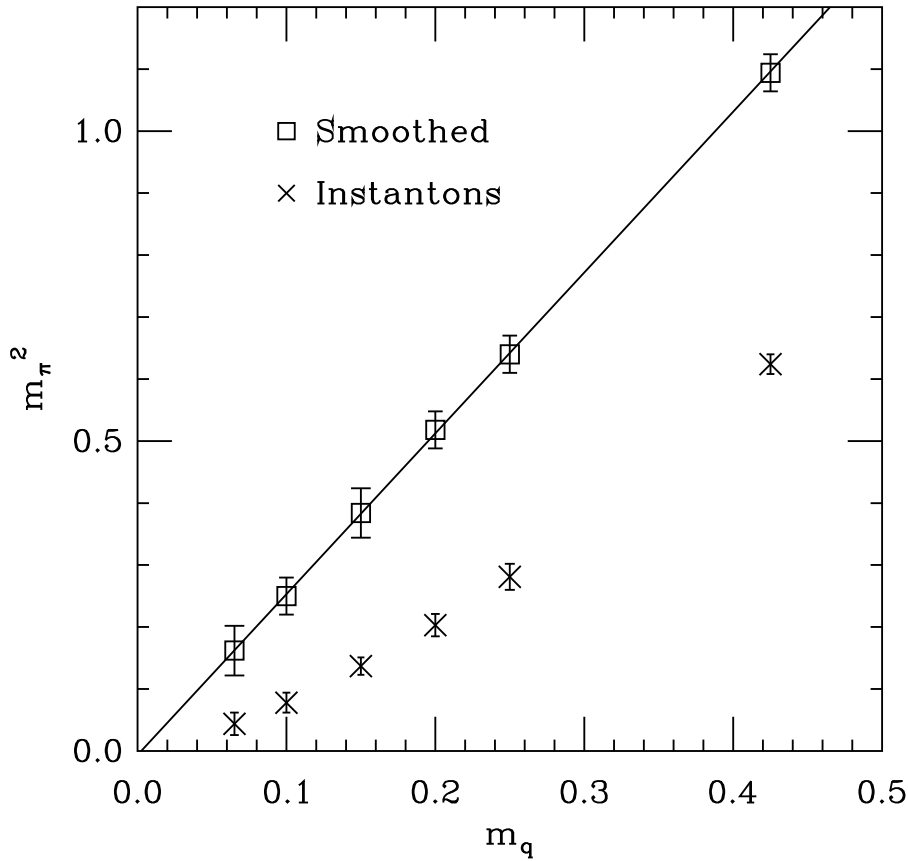}
\end{center}
\caption{The pion mass squared versus the bare quark mass on the 
smoothed (boxes) and instanton (crosses) ensemble.}
\label{fig:pi}
\end{figure}          
Fig.\ \ref{fig:pi} shows the pion mass sqared vs.\ the bare 
quark mass for both ensembles. On the smoothed ensemble $m_\pi^2 \propto
m_q$, as expected from PCAC.
This is clearly not the case on the instanton ensemble, where a 
best fit to the form $m_\pi^\lambda \propto m_q$ gives $\lambda=1.5$
which is between $\lambda=2$ and the free field value, $\lambda=1$.
We also note that since both ensembles are locally very smooth,
the additive mass renormalisations are very close to zero. Therefore,
it is also meaningful to compare hadron masses obtained at the same
bare quark mass on the two gauge ensembles. This comparison reveals
that the instanton ensemble typically produces much lighter hadrons
than the smoothed ensemble.

\subsection{Low Eigenmodes of the Dirac Operator}

In spite of its substantially improved
chiral properties, even the clover Dirac operator fails to
reproduce the correct physics of QCD on the instanton 
backgrounds. Instead, it yields a peculiar mixture of QCD and
free field theory. Is there a problem with the instanton liquid
model or the physics of these instanton lattices is still 
not close enough to the continuum? To answer this question 
we look at the low lying eigenmodes of the Dirac 
operator in more detail. 

\begin{figure}
\vspace{10mm}
\centerline{\ewxy{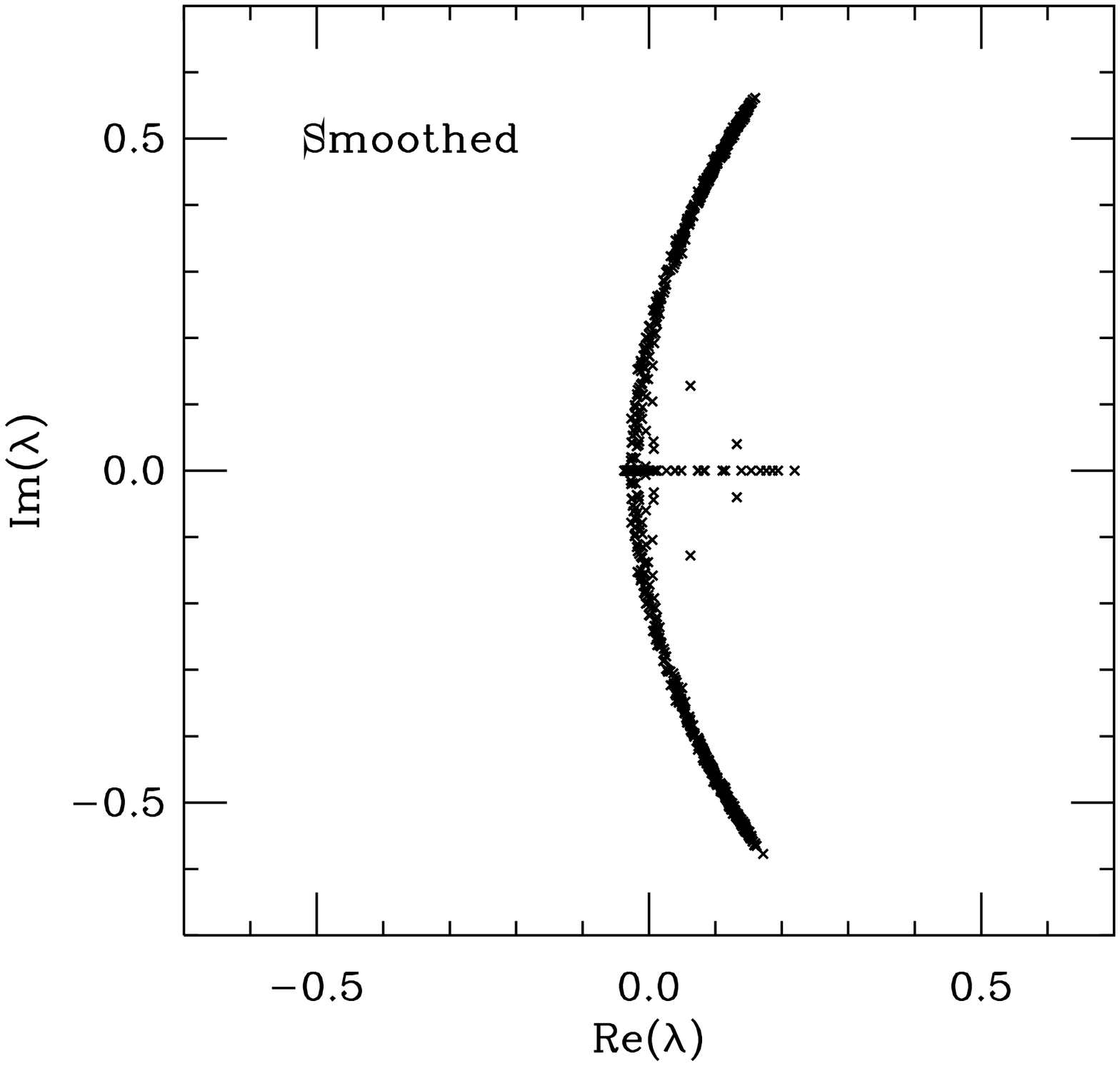}{80mm}
\ewxy{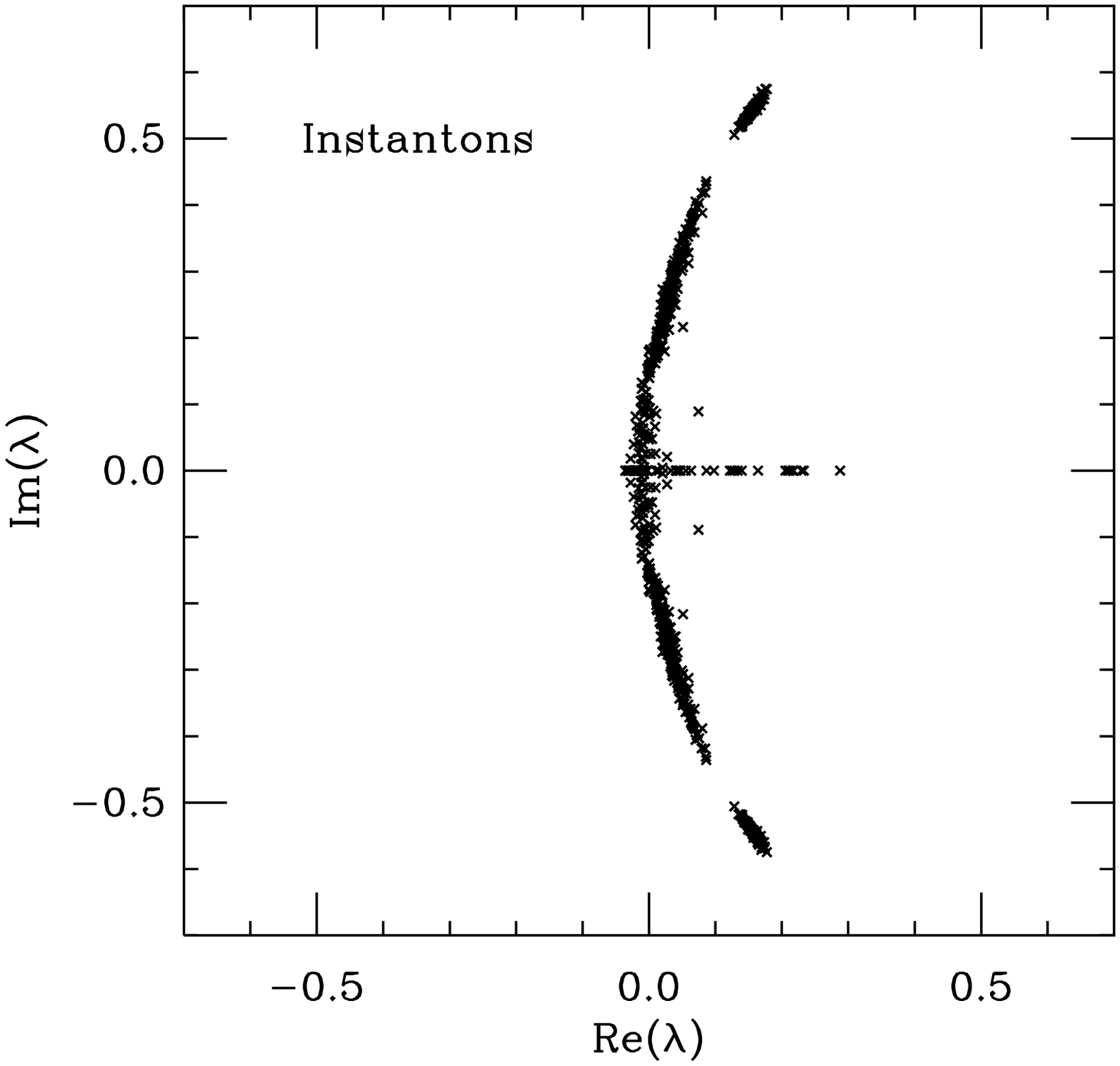}{80mm}}
\caption{The lowest 30 eigenvalues of the clover Dirac 
operator on the 28 smoothed and the corresponding 28 instanton 
gauge configurations.}
\label{fig:Dspec}
\end{figure}

In Fig.\ \ref{fig:Dspec} I plot the lowest 30 eigenvalues 
in the complex plane for the instanton and the smoothed 
ensemble. The eigenvalues of all 28 configurations are 
superimposed. The boundary condition is antiperiodic in the time
direction, periodic in all other directions.
In both cases most of the eigenvalues lie
very close to a circle passing through the origin. This is
a sign of the approximate chiral symmetry of the action.
The most striking difference between the two ensembles is 
the appearance of a gap at around Im$\lambda$=0.5 on the 
instanton ensemble. It is also instructive to 
compare the density of eigenvalues.
Since the low eigenvalues are almost imaginary I project 
them on the imaginary axis and plot the densities as a
function of the imaginary part of the eigenvalues.
This corresponds to the density of eigenvalues around 
zero in the continuum.

\begin{figure}[!htb]
\begin{center}
\vskip 10mm
\leavevmode
\epsfxsize=120mm
\epsfbox{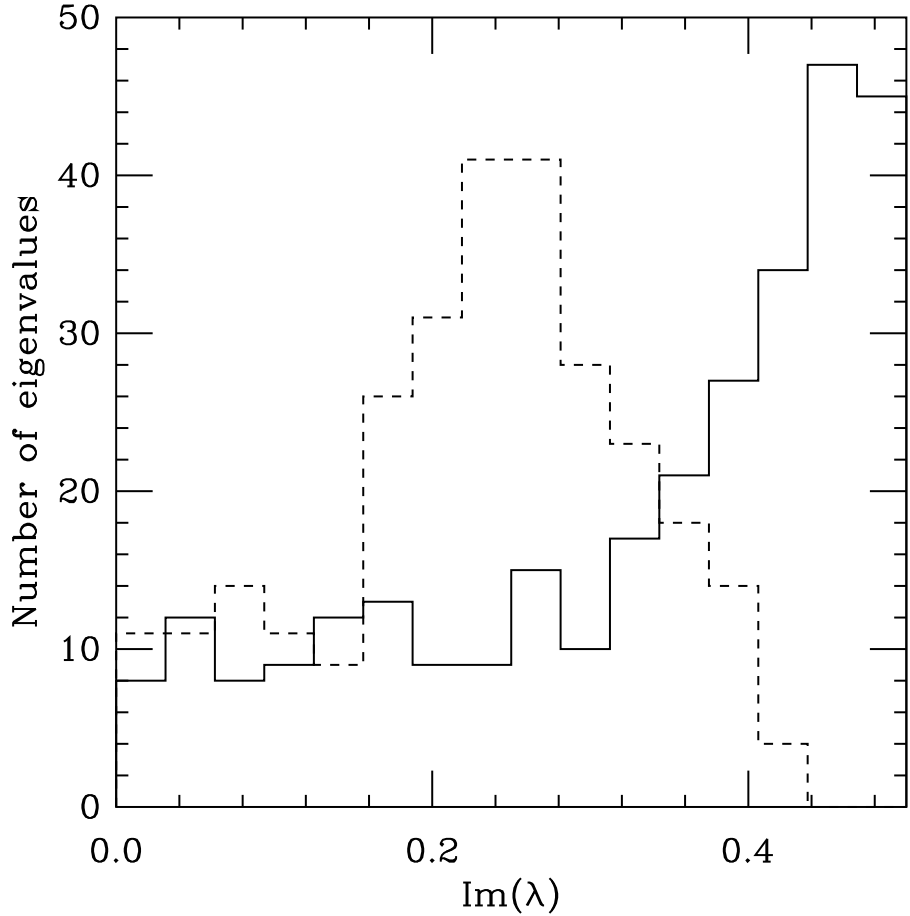}
\end{center}
\caption{The density of eigenvalues projected on the imaginary 
axis. The vertical axis shows directly the number of eigenvalues
in each bin. The solid line corresponds to the smoothed ensemble,
the dashed line to the instanton ensemble.}
\label{fig:ev_hist1}
\end{figure}          

In Fig.\ \ref{fig:ev_hist1} we can compare the densities 
corresponding to the two ensembles. The real modes have
been removed, these would show up as spikes at the origin.
The densities at zero seem to agree quite well. According 
to the Banks-Casher relation the chiral condensate is 
proportional to the density of modes around zero \cite{Banks-Casher}
(excluding exact zero modes, which do not contribute in the infinite
volume limit). This shows that on both the instanton and the 
smoothed ensemble chiral symmetry is broken and the value
of the chiral condensate is approximately the same. Apart 
from the vicinity of the origin, however, the two distributions
are quite different. On the instanton ensemble
there is a substantial ``piling up'' of modes above 
Im($\lambda)=0.2$ and then a ``thinning'' of modes at 0.5.

What is the reason of this huge difference between the two
distributions? The answer can be easily given by looking at
the eigenvectors of the Dirac operator. It turns out that
the quark density of the eigenvectors on the instanton ensemble 
corresponding to the peak of the distribution 
are much more delocalised than all the other eigenvectors occurring
in either ensemble. In fact, these modes are essentially 
free quark modes. 

This can be seen as follows. On a trivial gauge field 
configuration (all links are 1) with periodic bondary 
conditions in all directions, there are $4N_c$ trivial 
(constant) zero modes, this is the number of c-number
degrees of freedom corresponding to one fermion flavour.
In the presence of antiperiodic boundary condition in the 
time direction, the lowest eigenmodes are shifted away 
from zero, to
\begin{equation}
\mu_\pm = 1 - \cos\frac{\pi}{N_t} \pm i \sin\frac{\pi}{N_t} =
0.019 \pm 0.195 i,
\end{equation}
since in our case $N_t=16$. Both eigenvalues are $4N_c$-fold
degenerate. 

The mixing of the free field modes into a particular eigenmode
$\psi_\lambda$ can be characterised by $P_\pm(\lambda) =
\| P_\pm \psi_\lambda \|$, where $P_\pm$ is the projection of
the normalised eigenmode $\psi_\lambda$ onto the eigenspace 
corresponding to the eigenvalue $\mu_\pm$. This quantity is
not gauge invariant therefore I work in Landau gauge.
Any generic fermion mode on a non-trivial 
but locally smooth configuration will have a nonzero 
projection on the $\mu_\pm$ eigenspaces. The question is 
whether this is just an accidental mixing or the given
ensemble really contains close to free field modes. To decide
this, a good quantity to look at is $P(\lambda)=
P_+(\lambda)/P_-(\lambda)$. If the mixing is accidental, we
expect $P(\lambda)$ to fluctuate around 1, independently
of the corresponding eigenvalue $\lambda$. On the other
hand, if there are free field like modes on a given 
configuration then $P(\lambda)$ will increase substantially
when $\lambda$ approaches the free field eigenvalue $\mu_+$. 
 
\begin{figure}[!htb]
\begin{center}
\vskip 10mm
\leavevmode
\epsfxsize=120mm
\epsfbox{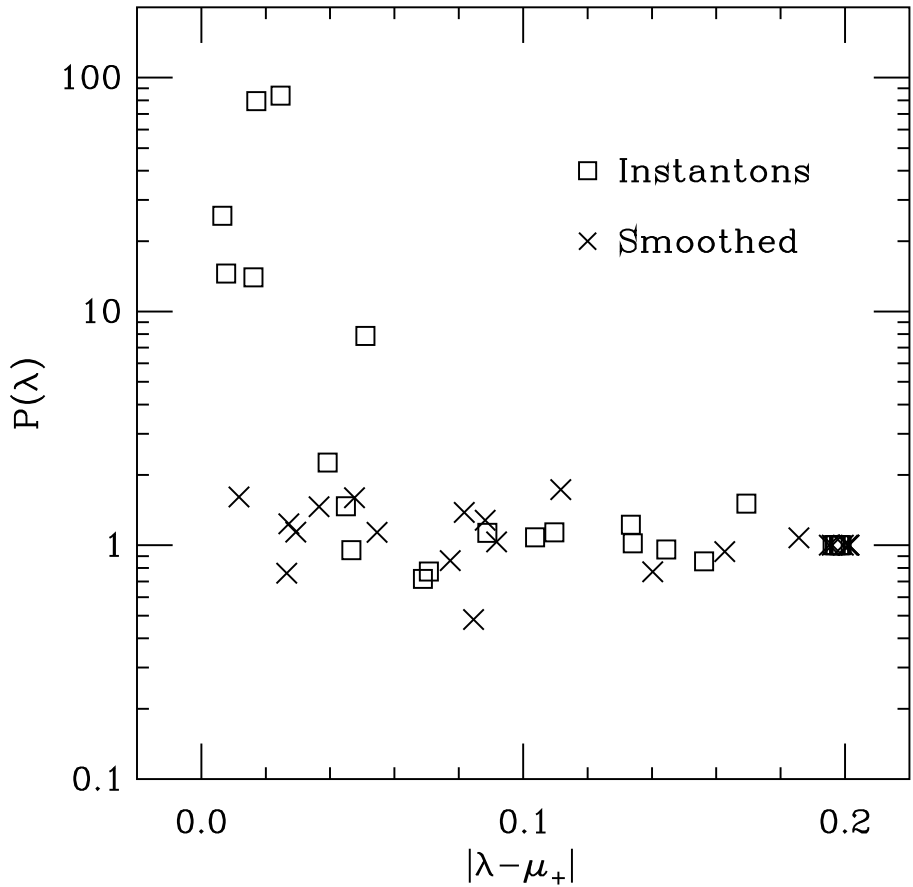}
\end{center}
\caption{$P(\lambda)=P_+(\lambda)/P_-(\lambda)$ versus the distance
of the eigenvalues $(\lambda)$ from the free field eigenmode
$\mu_+$ for the instantons only (boxes) and the smoothed 
(crosses) ensemble.}
\label{fig:proj}
\end{figure}          

In Fig.\ \ref{fig:proj} I plotted $P(\lambda)$ versus the
distance of the given eigenvalue from the free field mode
$\mu_+$. The Figure shows a random selection of eigenvalues 
with imaginary parts between 0.0 and 0.2.
In the smoothed ensemble $P(\lambda)$ 
fluctuates around 1 everywhere, there is no trace of the
free field modes. On the other hand, in the instanton ensemble,
the modes close to $\mu_+$ have a substantially larger 
projection on the $\mu_+$ eigenspace than on the $\mu_-$
subspace. In fact, these modes, close to $\mu_+$ have 
$\|P_+(\lambda)\| \approx 1$, so
they are essentially free field modes. A detailed study of
the chiral density $\psi_\lambda^\dagger \gamma_5 \psi_\lambda$
reveals that all the modes with $P(\lambda)\approx 1$ look like
mixtures of instanton zero modes with the density 
concentrated in several lumps. This is to be contrasted with
the chiral density of the modes with $P(\lambda) \gg 1$
that spreads almost homogenously over the whole lattice,
as expected of the lowest free field eigenmodes. 

Now we can understand why the instanton ensemble gave substantially
smaller hadron masses than the smoothed ensemble. The reason is
that the former had a large number of free quark modes. Even if
the zero mode zones of the two ensembles are similar, the
free field modes provide a very efficient way to propagate quarks
to large distances. This is why the instanton ensemble shows
a peculiar mixture of QCD and free field characteristics.
In the smoothed ensemble confinement completely eliminates the
freely propagating quarks.

\subsection{Suppression of the Free Quark Modes}

According to the instanton liquid model, instantons alone,
without confinement, can reproduce most of the properties of
the light hadrons. Our result shows that in the presence
of instantons only, without confinement, free quark modes 
provide a very effective way to propagate quarks and they
contaminate the hadron correlators producing unusually light
hadrons.

One possible way to suppress the free quark modes is to 
add to the instantons some locally very smooth background
that contains only long wavelength fluctuations. There has
to be enough long distance structure to suppress 
the free propagation of quarks but on the other hand
they have to be locally smooth enough not to distort the
instantons. Cycling is a very efficient way to create
such backgrounds. To this end I started with an 
ensemble of Wilson $\beta=2.4$ ($a=0.12$fm) $4^3\times8$
SU(2) gauge configurations, performed 4 cycling steps
and finally inverse blocked them to size $8^3\times16$.
This produced a set of lattices extremely smooth on the
few lattice spacing scale but containing the longest 
wavelength fluctuations characteristic to a 
$0.5^3\times1.0$fm$^4$ lattice. Their spatial size was a bit
smaller, their temporal size a bit larger than the confinement
scale. Moreover these configurations
had a small enough physical size that they almost never 
contained instantons. After superimposing these lattices on the
instanton configurations by simply multiplying the corresponding
links (in Landau gauge), I checked that the original 
topological charge of the instanton configurations changed 
very little. 

\begin{figure}
\vspace{10mm}
\centerline{\ewxy{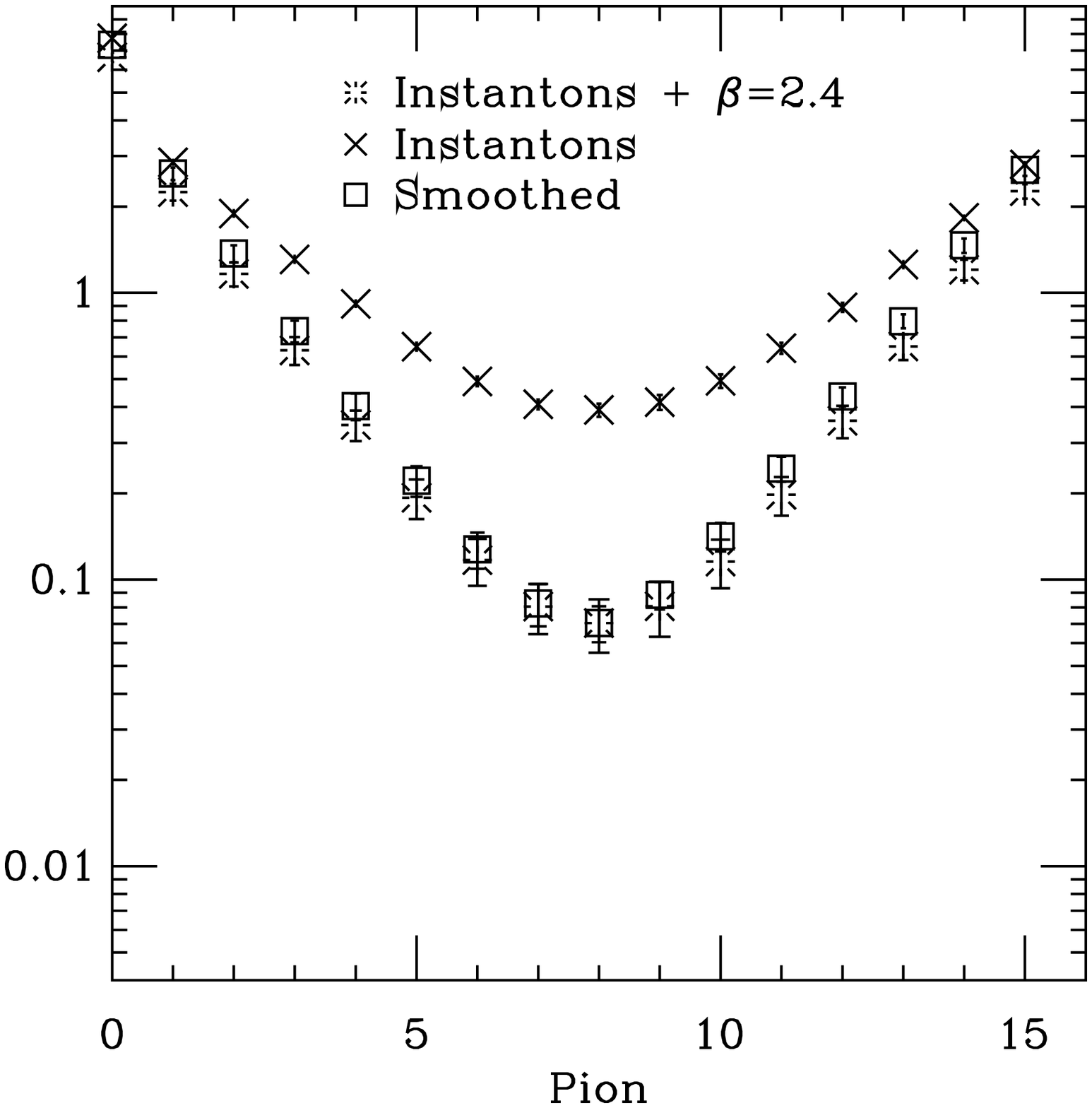}{80mm}
\ewxy{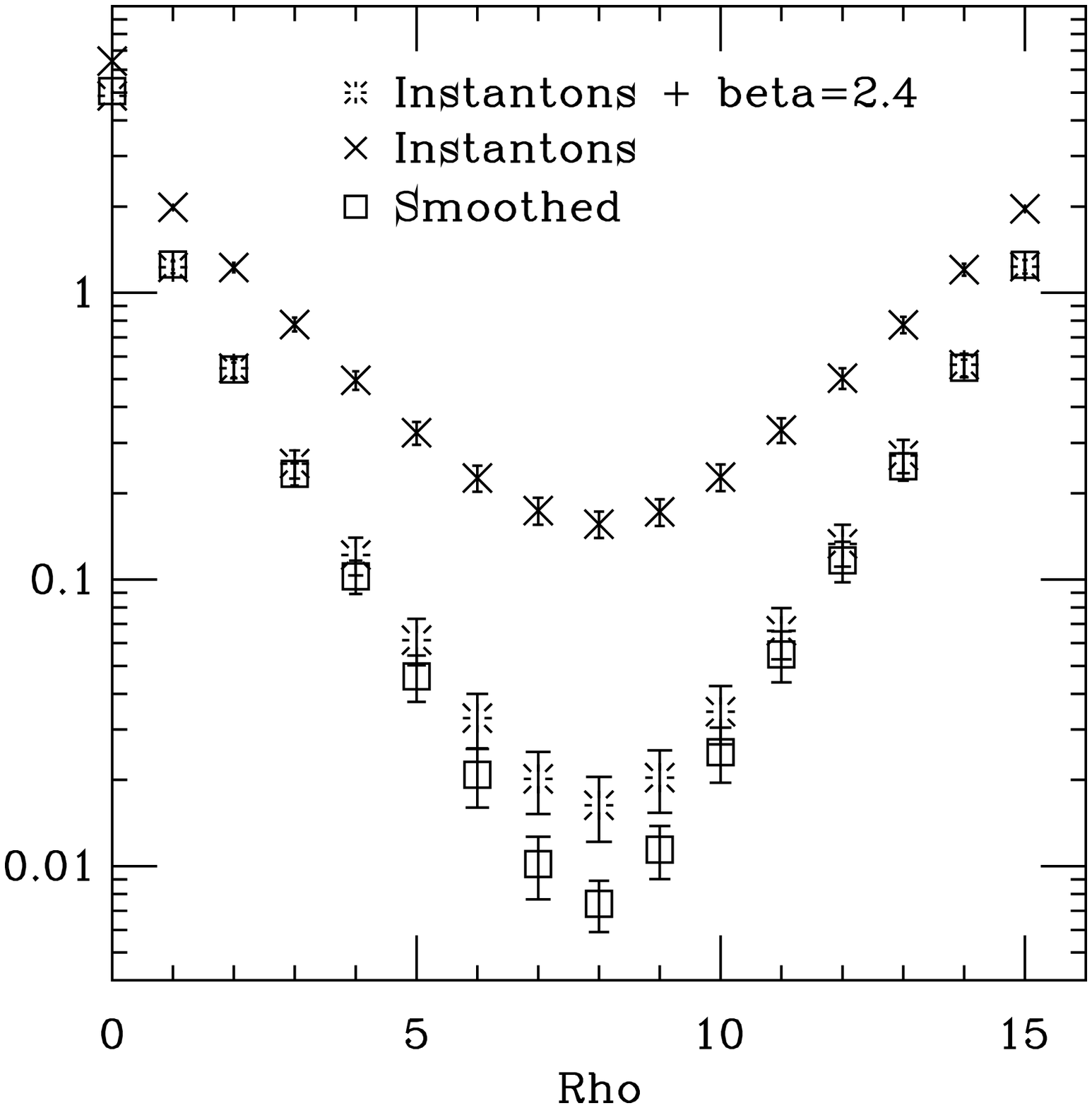}{80mm}}
\caption{Typical pion and rho correlators ($m_\pi/m_\rho=0.7$)
on the smoothed lattices (boxes), instanton configurations (crosses) 
and on the smooth background superimposed on the instantons
(bursts).}
\label{fig:pirho_b2.4}
\end{figure}

To show how the addition of the smooth background affects
the hadron correlators, in Fig.\ \ref{fig:pirho_b2.4} I plotted
typical pseudoscalar and vector correlators (at $m_\pi/m_\rho=0.7$).
The change is dramatic. While on the instanton configurations
the pion and the rho are very light (crosses), 
the addition of the smooth background
brings the correlators (bursts) very close to the physical ones
obtained on the smoothed lattices (boxes). The good agreement
of the physical correlators and the instanton plus smooth 
background correlators persists in the whole range of masses
where I tested it ($0.6<m_\pi/m_\rho<0.95$).

\begin{figure}
\vspace{10mm}
\centerline{\ewxy{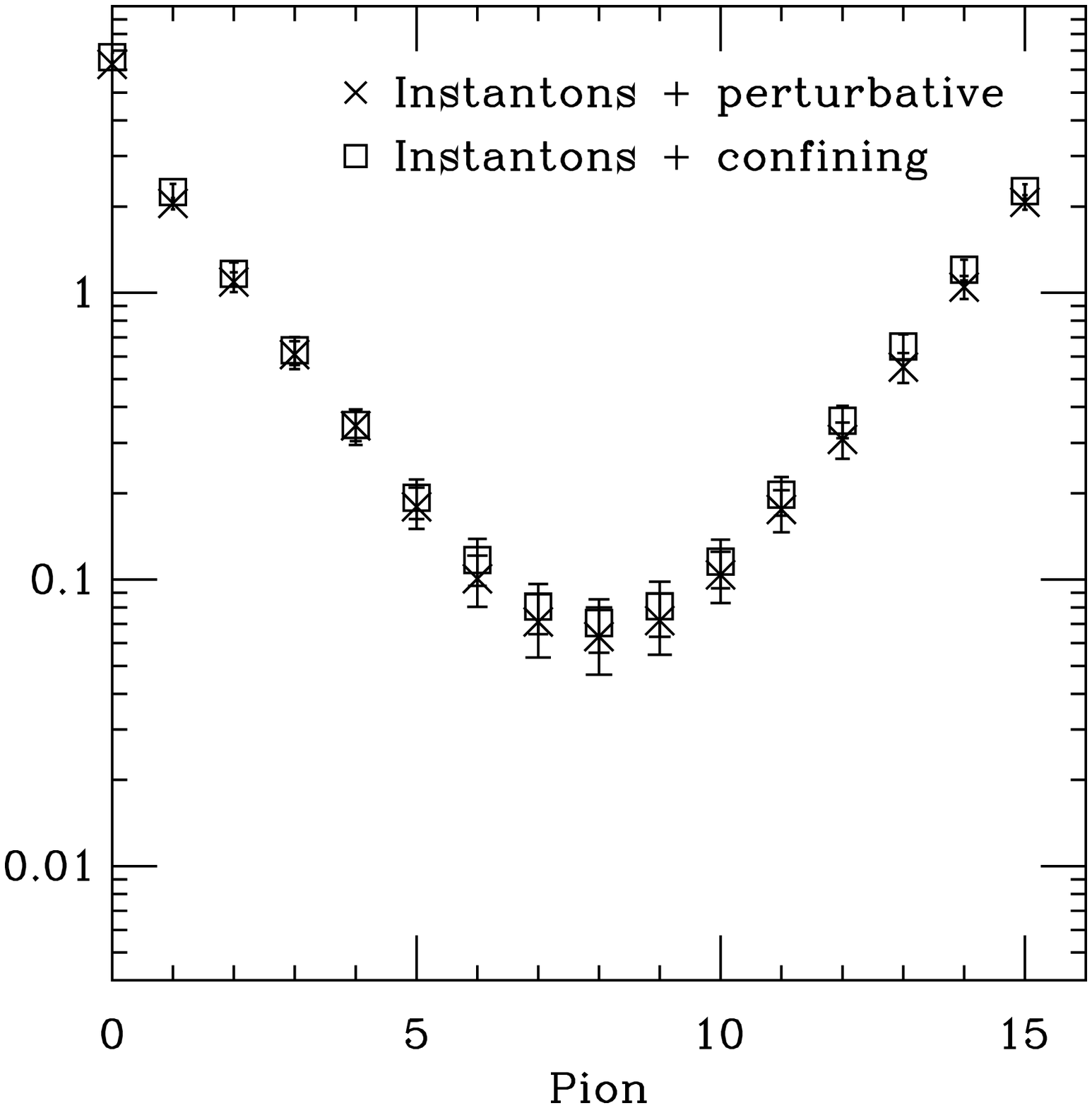}{80mm}
\ewxy{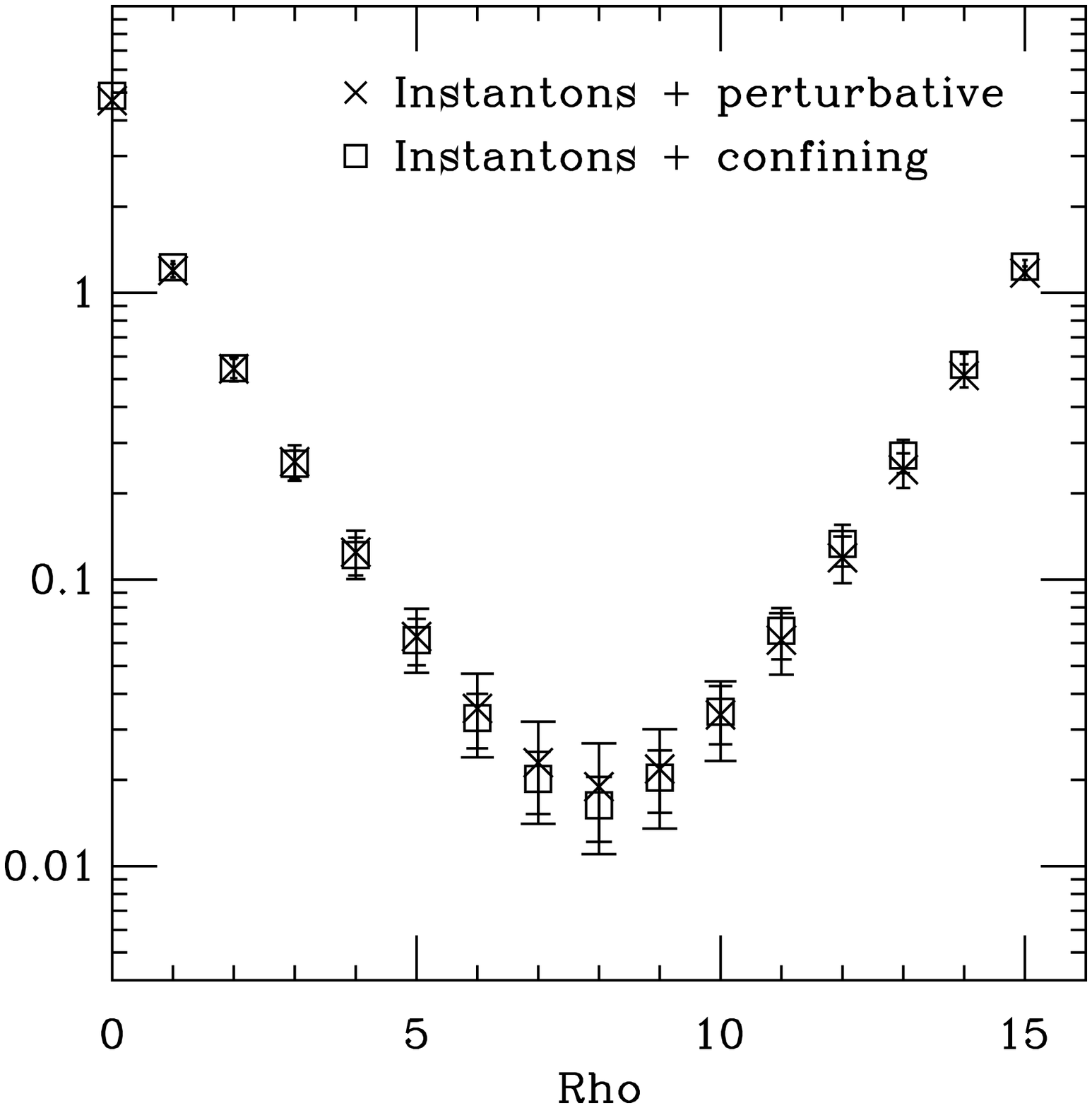}{80mm}}
\caption{Pion and rho correlators (at $m_\pi/m_\rho=0.7$)
on the instantons plus perturbative (crosses) and instantons
plus confining background (boxes).}
\label{fig:pirho_b3.0}
\end{figure}

It would be tempting to interpret this as a consequence of 
confinement; the superimposed smooth backgrounds still 
contained the longest scale fluctuations of the confining 
lattices and these removed the free field modes that 
contaminated the spectroscopy before. But is confinement
really needed for that? This can be easily tested by replacing
the superimposed backgrounds with ones of much smaller physical
size. The lattices that I used for this purpose were 
generated in exactly the same way as before, except 
that the starting $4^3\times 8$ lattices were produced at 
Wilson $\beta=3.0$. These lattices have a tiny physical size, 
they are expected to be completely perturbative.  
In Fig.\ \ref{fig:pirho_b3.0} I plotted again the pion and
rho correlators at $m_\pi/m_\rho=0.7$. Quite surprisingly,
the correlators have absolutely no dependence on the 
$\beta$ at which the background was created,
they are the same on the instantons plus confining and 
instantons plus perturbative configurations.

Apparently, there is still something nontrivial on these
physically very tiny perturbative lattices that has the same
effect on the correlators as a confining background. 
What could this be? The answer can be given by noting that
the $\beta \rightarrow \infty$ limit of Yang-Mills theory 
on the torus is not completely trivial because there is
not only one gauge field configuration with zero action. 
The set of flat (minimal action) gauge field configurations
can be characterised by the SU(2) conjugacy 
classes represented by four mutually commuting gauge
group elements corresponding to the (constant)  Polyakov loops
in the four perpendicular directions.\footnote{The 
general statement is that the gauge equivalence classes 
of flat connections on a connected manifold are in one-to-one 
correspondence with the conjugacy classes of representations 
of the first homotopy group of the base space in the gauge group 
\cite{Donaldson}.} These are essentially zero momentum
gauge field configurations, with a constant gauge potential 
(link), sometimes referred to as torons.
Their contribution to physical quantities is a finite 
size effect, and has been estimated in Ref.\ \cite{toron}.

To explicitly check that it is really the torons that 
affect the correlators so strongly, I extracted the ``toron
content'' of the perturbative background fields. This was
done in the following way. After Landau gauge fixing 
I averaged the Polyakov loops in all four directions. 
Since the Polyakov loops along any given direction were 
almost constant (up to small perturbative fluctuations), the averages
were close to being SU(2) elements. I projected the averages
back onto SU(2) and then distributed the average Polyakov
loop evenly among the links in the given direction.
This resulted in configurations with constant gauge fields
(links in any direction) that carried the average Polyakov
loops of the original perturbative configurations. I shall
refer to these lattices as ``toron'' configurations.

\begin{figure}
\vspace{10mm}
\centerline{\ewxy{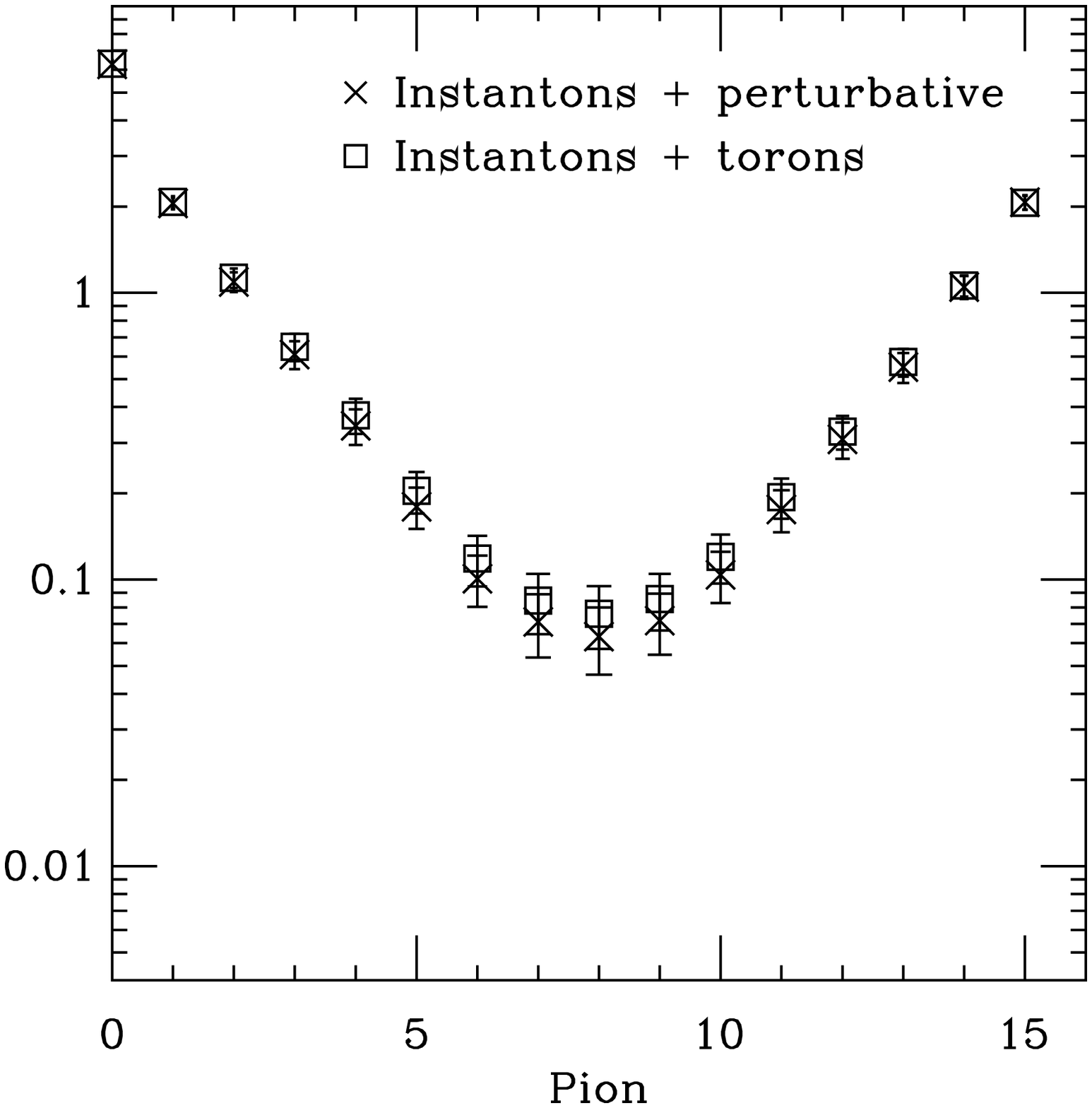}{80mm}
\ewxy{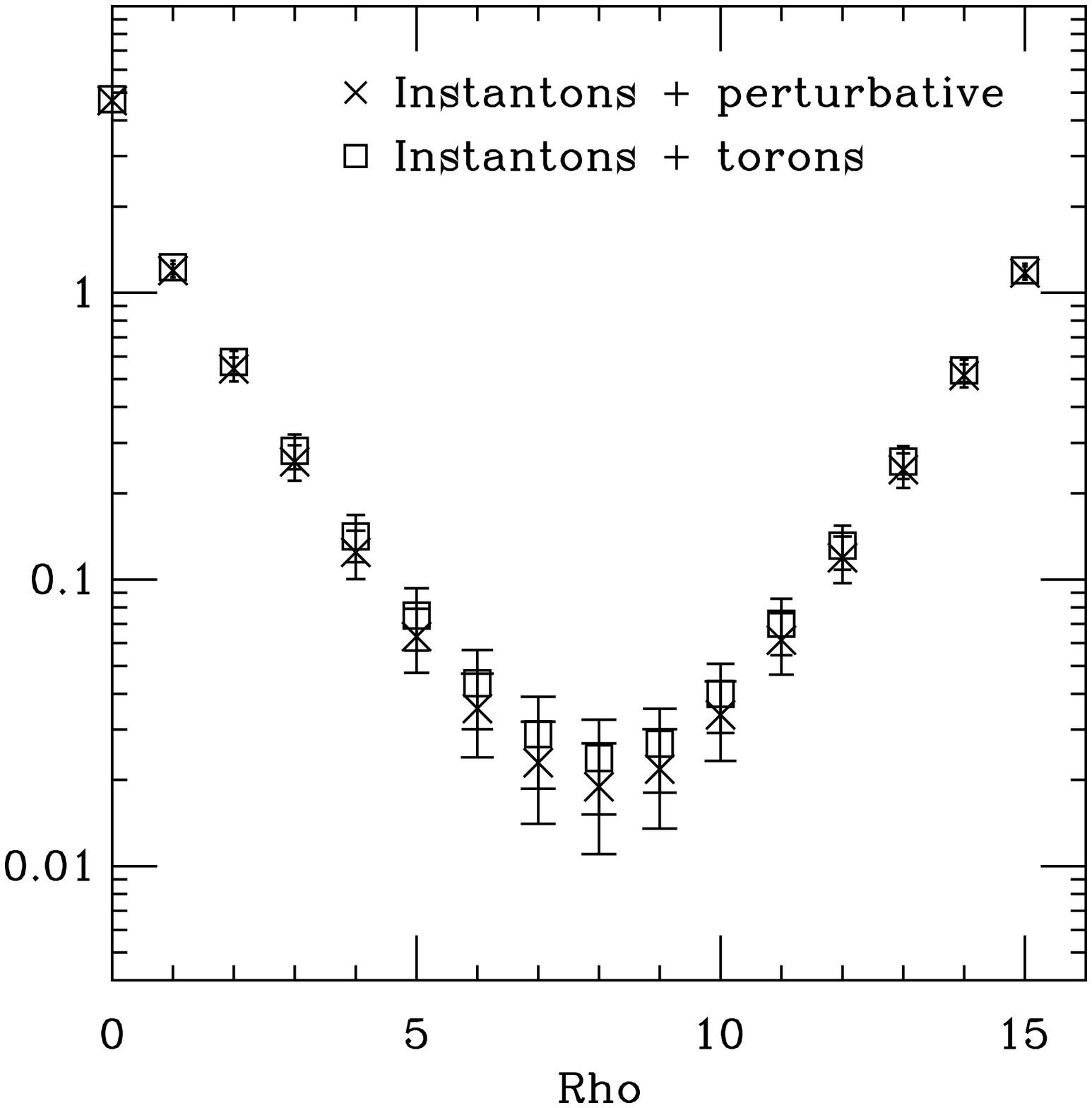}{80mm}}
\caption{Pion and rho correlators (at $m_\pi/m_\rho=0.7$)
on the instantons plus perturbative (crosses) and instantons
plus torons background (boxes).}
\label{fig:toron}
\end{figure}

We can now compare the correlators in the instantons plus
toron background and the instantons plus the full 
perturbative background. The result --- again in the 
typical case of $m_\pi/m_\rho=0.7$ --- is shown in
Fig.\ \ref{fig:toron}. The instantons and the torons
together fully reproduce the correlators, demonstrating that
the most important fluctuations on the perturbative lattices are
indeed the torons. 

\begin{figure}[!htb]
\begin{center}
\vskip 10mm
\leavevmode
\epsfxsize=120mm
\epsfbox{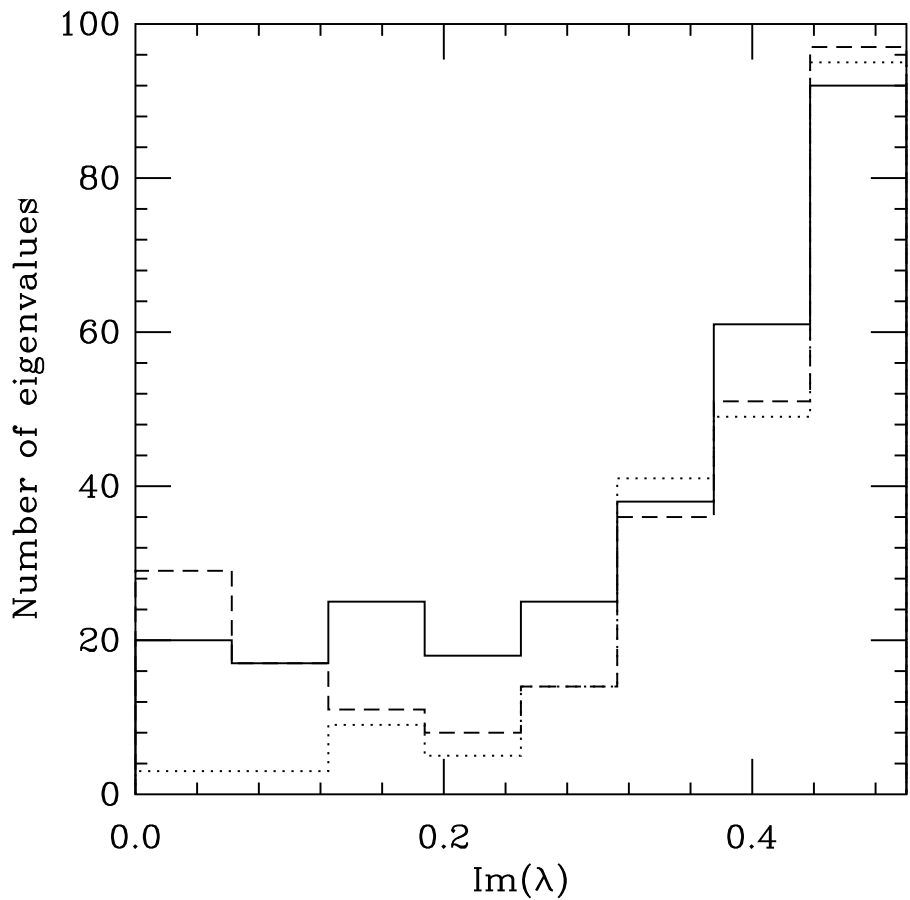}
\end{center}
\caption{The density of eigenvalues of the Dirac operator
on the smoothed ensemble (solid line), the perturbative 
ones (dotted line) and the instantons plus perturbative 
background (dashed line).} 
\label{fig:bc_hist}
\end{figure} 

It is also quite instructive to compare the distribution 
of the lowest eigenvalues of the Dirac operator on the 
different types of gauge configurations. Fig.\ \ref{fig:bc_hist}
shows the densities on the smoothed lattices
(solid lines), the perturbative ones (dotted line) and the 
lattices containing both the instantons and the perturbative 
backgrounds (dashed line). 
The three distributions agree quite well away from zero.
However around zero, the perturbative ensemble produces
only a very low density (compatible with zero), while
the two other ensembles yield qualitatively similar
densities, both nonzero. This demonstrates very 
clearly that it is the instantons that are responsible for
creating a nonzero eigenvalue density around zero which
through the Banks-Casher relation \cite{Banks-Casher} implies
the spontaneous breaking of chiral symmetry. We also note 
that the peak corresponding to the free field modes
(see Fig.\ \ref{fig:ev_hist1}) has been completely eliminated
by the perturbative backgrounds. This again can be
attributed to the torons, as can be checked by looking at the
eigenvalue distribution on the toron lattices (not shown)
which is very similar to the dotted line in Fig.\ \ref{fig:bc_hist}.

\section{Conclusions}
\label{se:con}

In the present paper I addressed the question of how much
of QCD in the chiral limit is reproduced by restricting the
fluctuations of the gauge field to instantons only. 
I reconstructed the instanton content of smoothed Monte
Carlo lattice configurations and compared hadron spectroscopy
on this instanton ensemble to the spectroscopy on  the original
``physical'' smoothed configurations. The fermion action used
for the spectroscopy was a ``chirally optimised'' clover action.
I explicitly studied the fermion zero modes in the presence
of an instanton and also the mixing of the zero modes 
corresponding to an instanton and an antiinstanton. I found that
in these simple cases the optimised action described 
the continuum features of the zero mode zone rather well. 
In spite of this, the optimised action still failed to
reproduce the physical hadron spectrum. It yielded anomalously
light hadrons with a mixture of QCD and free-field like
features. 

A closer look at the spectrum of the Dirac operator revealed 
that on the instanton ensemble it had a large number of
free quark modes. These were absent on the physical smoothed
configurations. They provided a very efficient way of 
propagating quarks to large distances thereby substantially
reduced the hadron masses and contaminated the spectroscopy. 
Superimposing an ensemble of very smooth, essentially 
perturbative gauge field fluctuations on the instantons,
was enough to eliminate the free quark modes
and to restore the physical hadron correlators. 
By comparing the density of low eigenvalues of the Dirac 
operator I also obtained direct evidence that it is the instantons 
that are responsible for creating a nonzero density 
of modes around zero and by the Banks-Casher relation 
also for chiral symmetry breaking.

The important fluctuations on the superimposed perturbative
lattices, needed to recover the physical hadron correlators,
turned out to be zero momentum (constant link or
gauge potential) configurations, torons. I would like
to emphasise here that this toron ensemble was a perturbative
one, produced at large $\beta$ i.e. small physical lattice
size. The meson correlators appeared to be largely independent
of the $\beta$ at which the superimposed background was
created, as long as it was above the finite temperature
phase transition on the given lattice size. 

It seems rather surprising that such ``mild'' gauge field
fluctuations as the torons, can have so profound an impact
on physical quantities. After all, torons exist only on
the torus, they are absent in an infinite space-time
and thus constitute only finite size effects. On the other
hand we know that the instantons by themselves do not
confine \cite{noconf,spect}. This means that the instanton
ensemble does not produce a mass gap and it is expected 
to exhibit much stronger finite size effects than QCD, from 
which the instantons have been extracted. In fact, the
free field modes that are eliminated by the torons are
also absent in the infinite volume limit, since they
are not normalisable. In free field theory the density
of modes around zero is proportional to the linear
size of the box $V^{1/4}$, whereas the density of instanton
related modes grows proportionally to the volume. It is
thus not inconceivable that in increasingly bigger volumes,
although the torons have less and less influence, at the same time
they become less and less needed since the relative importance
of the free field modes also dies out. The torons might
be needed only for compensating the unnaturally large 
finite size effects due to the absence of confinement in 
the instanton ensemble. 

What is the picture emerging from this lattice study 
concerning the instanton liquid model?
The QCD vacuum contains an ensemble of instantons that 
is generated by the non-perturbative gauge field dynamics.
The instantons by themselves break chiral symmetry but
hadron correlators in the instanton backgrounds are strongly
contaminated by freely propagating quarks. This yields
anomalously light mesons and a small splitting between the
pseudoscalar and the vector channel. The free quarks can
be eliminated and the physical hadron correlators can be
restored by superimposing a perturbative ensemble of
zero momentum gauge configurations (torons) on the instantons.
This suggests that the contamination by free quarks 
is most likely a finite volume effect which, in QCD, is
completely suppressed by confinement.

More evidence in favour of this scenario could be gathered by
studying the volume dependence of these effects on larger volumes.
Another important question, not answered by the 
present study is why it is exactly the
perturbative toron ensemble that is needed to restore the
physical meson correlators. It would be also 
very desirable to extend this investigation to a larger
class of observables, in particular baryon correlators,
for which the SU(3) case need be considered. Finally,
for more precise tests a better chiral action, such as 
the overlap \cite{overlap} could be used.

\section*{Appendix}

In the Appendix, for reference, I collected some properties 
of Wilson type lattice Dirac operators that were used in the main
body of the paper. Some of these results can be found in
the literature, in particular in Refs.\ \cite{IIY,G-Hip,Leutwyler}. 

I assume that the Dirac operator satisfies 
$\gamma_5$ Hermiticity, i.e.\
\begin{equation}
  D^\dagger = \gamma_5 \; D \; \gamma_5,
       \label{eq:g5}
\end{equation}  

\begin{statement}
Let $D$ be a $\gamma_5$-Hermitian Dirac operator with no
degeneracy in its spectrum, i.e.\ having as many different
eigenvalues as the dimension of the space it acts on.
Let $\lambda$, and $\mu$ be two eigenvalues of $D$ with the 
corresponding eigenvectors being $\psi_\lambda$ and $\psi_\mu$.
Then $\psi_\mu^\dagger \gamma_5 \psi_\lambda \neq 0$ if and only if
$\mu^\star = \lambda$.
\end{statement}
{\em Proof:} A simple consequence of $\gamma_5$ Hermiticity is that
\begin{equation}
  \mu^\star \, \psi_\mu^\dagger \gamma_5 \psi_\lambda \; = \; 
  \psi_\mu^\dagger D^\dagger \gamma_5 \psi_\lambda \; = \;
  \psi_\mu^\dagger \gamma_5         D \psi_\lambda \; = \;
  \lambda \, \psi_\mu^\dagger \gamma_5 \psi_\lambda 
\end{equation}
which implies
\begin{equation}
  (\mu^\star - \lambda) \, 
  \psi_\mu^\dagger \gamma_5 \psi_\lambda = 0.
     \label{eq:g5pr}
\end{equation}
It immediately follows that if $\psi_\mu^\dagger \gamma_5 
\psi_\lambda \neq 0$ then $\mu^\star = \lambda$. 
To prove the converse we first note
that since $D$ has as many different eigenvalues as the
dimension of the space it acts on, its eigenvectors form
a basis. We have already seen that the only eigenvector
on which $\gamma_5 \psi_\lambda$ can have a nonzero projection,
is $\psi_{\lambda^\star}$. If this projection were also zero,
then $\gamma_5 \psi_\lambda$ would be orthogonal
to all the vectors in a complete set and consequently it
would be zero. This is impossible since $\gamma_5$ has a trivial
kernel.

For proving that $\psi_\lambda^\dagger \gamma_5 \psi_{\lambda^\star}
\neq 0$ it was essential that the eigenvalues of $D$ be 
non-degenerate and its eigenvectors form
a complete set. Generally this is expected to be the case unless
the gauge field is fine tuned. There is however an important
exception. When topology --- as seen by the fermions --- changes
through a smooth deformation of the gauge fields, two opposite
chirality real eigenvalues collide and leave the real axis as a 
complex conjugate pair. It can be shown that when the two 
eigenvalues coincide, the corresponding eigenspace is 
one-dimensional, the eigenvector has zero chirality, 
$D$ does not possess a complete set of eigenvectors and
therefore has no spectral decomposition.

A simple consequence of Statement 1 is that if
$D$ has a complete non-degenerate spectrum, then the chirality
of an eigenvector, $\psi_\lambda^\dagger \gamma_5 \psi_\lambda$
is nonzero if and only if the corresponding eigenvalue, $\lambda$,
is real.
 
In the remainder of the Appendix I shall prove the following
property of Wilson type SU(2) lattice Dirac operators.
\begin{statement}
Let $\psi$ and $\chi$ be real eigenvectors of two (not necessarily
the same) SU(2) lattice Dirac operators such that the corresponding
eigenvalues and their complex conjugates are non-degenerate.
It is then possible to choose the phases of $\psi$ and $\chi$ 
such $\psi^\dagger D \psi$, $\psi^\dagger D \chi$, $\chi^\dagger D \psi$, 
and $\chi^\dagger D \chi$ are all real for any (third) lattice 
Dirac operator $D$. 
\end{statement}
{\em Proof:} Besides $\gamma_5$ Hermiticity, this property depends
on an additional symmetry of the Dirac operator relating $D$ to
its complex conjugate \cite{Leutwyler}. 
This is specific to the SU(2) case. Let 
$C$ be the charge conjugation operator and $K=C^{-1}\otimes\tau_2$,
where $\tau_2$ is a Pauli matrix acting in colour space. Using
that  
\begin{equation}
 C^{-1} \, \gamma_\mu \, C \; = \; -\gamma_\mu^T \; = \; 
 -\gamma_\mu^\star,
\end{equation}
(where the last equality follows because we work in a 
representation with all $\gamma_\mu$'s Hermitian), and
that for any $U\in SU(2)$ $\tau_2 U \tau_2 = U^\star$,
it is not hard to prove that
\begin{equation}
  (K \gamma_5)^\dagger \, D \, K\gamma_5 \; = \; D^\star.
      \label{eq:dstar}
\end{equation} 
Now if $\psi_\lambda$ is an eigenvector of $D$ with eigenvalue
$\lambda$ then 
\begin{equation}
   D K\gamma_5 \psi_\lambda^\star \; = \; 
   K\gamma_5 D^\star \psi_\lambda^\star \; = \;
   \lambda^\star K\gamma_5 \psi_\lambda^\star,
\end{equation}
which means that $K\gamma_5 \psi_\lambda^\star$ is an eigenvector
of $D$ with eigenvalue $\lambda^\star$. Due to the non-degeneracy of
the eigenvalues and that $(K\gamma_5)^\dagger K\gamma_5=1$, it follows
that 
\begin{equation}
  K\gamma_5 \psi_\lambda^\star = e^{i\phi} \psi_{\lambda^\star}.
    \label{eq:phase}
\end{equation}
In the special case when $\lambda$ is real, the phase choice
$e^{i\phi/2}\psi_\lambda$ eliminates the extra phase factor
on the r.h.s. of eq.\ (\ref{eq:phase}). 

Now writing $(\psi^\dagger D \chi)^\star = 
(\psi^\star)^\dagger D^\star \chi^\star$, and making use of
eqs. (\ref{eq:dstar}) and (\ref{eq:phase}) with the above
phase choice, the non-diagonal matrix elements can be easily
seen to be real. The same argument also shows that the diagonal
matrix elements are automatically real, regardless of the 
phase choice. This completes the proof.

We would like to note that while any Wilson type
Dirac operator has a spectrum symmetric with respect to the real
axis (this follows from $\gamma_5$ Hermiticity), in general
there is no simple relation between the eigenvectors corresponding
to complex conjugate eigenvalues. The only exception is
the SU(2) case, when they are related by eq.\ (\ref{eq:phase}).

\section*{Acknowledgements}

I thank Pierre van Baal for discussions,
Tom DeGrand and Anna Hasenfratz for correspondence.
The computations were performed using a computer code
based on the MILC \cite{MILC} code and the ARPACK \cite{ARPACK}
package. I thank the MILC Collaboration and the
authors of the ARPACK package for making their code publicly
available. This work has been supported by FOM.


\begin{thebibliography}{99}

\bibitem{Shuryak}
T.~Sch\"afer, and E.V.~Shuryak, Rev.\ Mod.\ Phys.\ {\bf 70} 
(1998) 323.

\bibitem{Diakonov} 
D.I.~Diakonov, and V.Yu.~Petrov, Nucl.\ Phys.\ {\bf B245} (1984) 259. 

\bibitem{everybody}
Ph.~de~Forcrand, M. Garc\'{\i}a P\'erez, and I.-O.~Stamatescu,
Nucl.\ Phys.\ {\bf B499} (1997) 409; 
C.~Michael, and P.S.~Spencer, Phys.\ Rev.\ {\bf D52} (1995) 4691; 
D.A.~Smith, and M.J.~Teper, Phys.\ Rev.\ {\bf D58} (1998) 014505;
B.~Alles, M.~Campostrini, A.~Di~Giacomo, Y.~Gunduc, E.~Vicari,
Phys.\ Rev.\ {\bf D48} (1993) 2284; 
A.\ Hasenfratz, and C.~Nieter, Phys.\ Lett.\ {\bf B439} (1998) 366. 

\bibitem{indirect}
R.G.~Edwards, U.M.~Heller, J.~Kiskis, and R.~Narayanan, 
Phys.\ Rev.\ Lett.\ {\bf 82} (1999) 4188;  R.G.~Edwards, 
U.M.~Heller, J.~Kiskis, and R.~Narayanan, Chiral condensate 
in the deconfined phase of quenched gauge theories, hep-lat/9919941.

\bibitem{For_ch}
Ph.~de~Forcrand, M.~Garcia~Perez, J.E.~Hetrick, E.~Laermann, 
J.F.~Lagae, and I.O.~Stamatescu, Nucl.\ Phys.\ Proc.\ Suppl.\
{\bf 73} (1999) 578.

\bibitem{Negele}
M.C.~Chu, J.M.~Grandy, S.~Huang, and J.W.~Negele, Phys.\ Rev.\
{\bf D49} (1994) 6039.

\bibitem{spect}
T.~DeGrand, A.~Hasenfratz, and  T.G.~Kov\'acs, Phys.\ Lett.\ {\bf  B420}
(1998) 97.
 
\bibitem{exceptional}
T.~DeGrand, A.~Hasenfratz, and  T.G.~Kov\'acs, Nucl.\ Phys.\ {\bf B547}
(1999) 259, also available as hep-lat/9810061v2.

\bibitem{scaling}
M.~Stephenson, C.~DeTar, T.~DeGrand, and A.~Hasenfratz,
Scaling and eigenmode tests of the improved fat clover action,
hep-lat/9910023.

\bibitem{Ni-Ni}
H.B.~Nielsen, and M.~Ninomiya, Phys.\ Lett.\ {\bf 105B} (1981) 219. 

\bibitem{HP}
P.~Hasenfratz, V.~Laliena, and F.~Niedermayer, Phys.\ Lett.\ 
{\bf B427} (1998) 125.

\bibitem{singleI}
R.G.~Edwards, U.M.~Heller, and R.~Narayanan, Nucl.\ Phys.\
{\bf B522} (1988) 285. 

\bibitem{Boulder1}
T.~DeGrand, A.~Hasenfratz, and  T.G.~Kov\'acs, Nucl.\ Phys.\ {\bf B505}
(1997) 417.

\bibitem{Ilgenfritz}
E.M.~Ilgenfritz, and S.~Thurner, Correlated instanton orientations
in the SU(2) Yang-Mills vacuum and pair formation in the 
deconfined phase, hep-lat/9810010.
   
\bibitem{Banks-Casher}
T.~Banks, and A.~Casher, Nucl.\ Phys.\ {\bf B167} (1980) 215.

\bibitem{Donaldson}
S.K.~Donaldson, and P.B.~Kronheimer, The geometry of four-manifolds,
Clarendon Press, Oxford, 1990, pp.\ 49.

\bibitem{toron}
A.~Coste, A.~Gonzalez-Arroyo, J.~Jurkiewicz, and 
C.P.~Korthals~Altes, Nucl.\ Phys.\ {\bf B262} (1985) 67.

\bibitem{noconf}
D.I.~Diakonov, V.Yu.~Petrov, and P.V.~Pobylitsa, Phys.\ Lett.\
{\bf B226} (1989) 372.

\bibitem{overlap}
H.~Neuberger, Chiral fermions on the lattice, hep-lat/9909042,
and references therein.

\bibitem{IIY}
S.~Itoh, Y.~Iwasaki, and T.~Yoshie, Phys.\ Rev.\ {\bf D36} (1987)
527.

\bibitem{G-Hip}
C.R.~Gattringer, I.~Hip, and C.B.~Lang, Nucl.\ Phys.\ 
{\bf B508} (1997) 329.

\bibitem{Leutwyler}
H.~Leutwyler, and A.~Smilga, Phys.\ Rev.\ {\bf D46} (1992) 5607.

\bibitem{MILC}
The MILC code is available at
ftp://ftp.physics.utah.edu/pub/milc/freeHEP/.

\bibitem{ARPACK}
D.C.~Sorensen, SIAM J.\ Matrix Analysis and Applications,
{\bf 13(1)} (1992) 357; the package 
is available at http://www.caam.rice.edu/software/ARPACK/.

\end{thebibliography}
\end{document}